\documentclass[usenatbib]{mn2e}
\usepackage{times}
\usepackage{epsfig}
\usepackage{amsmath}
\usepackage{amssymb}
\title[Overconsumption in satellite galaxies]{Overconsumption, outflows and the quenching of satellite galaxies}

\author[S. L. McGee et al.]{Sean L. McGee$^{1}$\thanks{Email: mcgee@strw.leidenuniv.nl},  
Richard G. Bower$^{2}$, 
Michael L. Balogh$^{1,3}$
 \\
$^{1}$ Leiden Observatory, Leiden University, PO Box 9513, 2300 RA 
Leiden, The Netherlands\\ 
$^{2}$Department of Physics, University of Durham, Durham, UK, DH1 3LE\\
$^{3}$Department of Physics and Astronomy, 
University of Waterloo, Waterloo, Ontario, N2L 3G1, Canada\\
}

\date{\today}
\def\LCDM{$\Lambda$CDM$~$}

\def\Mdot{M$_\odot$}
\def\Msun{M$_\odot$}
\def\Mstel{M$_{stel}$}

\def\kmsmpc{\>{\rm km}\,{\rm s}^{-1}\,{\rm Mpc}^{-1}}
\def\galex{{\it GALEX}\ }

\def\beq{\begin{equation}}
\def\eeq{\end{equation}}

\def\dotMin{\dot{M}_{in}}
\def\dotMres{\dot{M}_{res}}
\def\dotMstar{\dot{M}_{stel}}
\def\dotMout{\dot{M}_{out}}

\begin{document}

\maketitle
\begin{abstract}

The baryon cycle of galaxies is a dynamic process involving the intake, consumption and ejection of vast quantities of gas. In contrast, the conventional picture of satellite galaxies has them methodically turning a large gas reservoir into stars until this reservoir is forcibly removed due to external ram pressure. This picture needs revision. Our modern understanding of the baryon cycle suggests that in some regimes the simple interruption of the fresh gas supply may quench satellite galaxies long before stripping events occur, a process we call overconsumption. We compile measurements from the literature of observed satellite quenching times at a range of redshifts to determine if satellites are principally quenched through orbit-based gas stripping events -- either direct stripping of the disk (ram pressure stripping) or the extended gas halo (strangulation) -- or from internally-driven star formation outflows via overconsumption. The observed timescales show significant deviation from the evolution expected for gas stripping mechanisms and suggest that either ram pressure stripping is much more efficient at high redshift, or that secular outflows quench satellites before orbit-based stripping occurs. Given the strong redshift evolution of star formation rates, at high redshift (z $>$ 1.5) even moderate outflow rates will lead to extremely short quenching times with the expectation that such satellites will be quenched almost immediately following the cessation of cosmological inflow, regardless of stripping events. Observations of high redshift satellites give an indirect but sensitive measure of the outflow rate with current measurements suggesting that outflows are no larger than 2.5 times the star formation rate for galaxies with a stellar mass of 10$^{10.5}\ $\Msun. 
\end{abstract}


\section{Introduction}

The conventional picture of the interaction between galaxies and their environments is based on the idea that galaxies enter dense environments with a reservoir of gas.  The removal of this reservoir, either from the stellar disk (ram pressure stripping), or the galaxy's halo (strangulation), leads to a decline in the star formation rate. This view is out of date with our understanding of the dynamic baryon cycle in galaxies. Cosmological simulations, as well as semi-analytic models, show that galaxies grow as a result of gas infall from surrounding filaments. At least at high redshift, rapid star formation is fed by a supply of fresh, infalling gas which dominates over the consumption of any reservoir \citep[eg.,][]{papovich_2011_short}. Taking this into account leads to a significant update to our picture of how galaxies interact with their environment. If this scenario is correct, simply removing the supply of fresh material will have a dramatic effect on the properties of satellite galaxies, without the need for gas stripping processes. In this Letter, we use a compilation of observed satellite star formation timescales at a range of redshifts, to examine the mechanisms which quench satellites and the insights this gives to the baryon cycle of all galaxies. 

In recent years, many authors have combined direct observations of satellite quenched fractions with cosmological infall rates of galaxies to estimate the time it takes a satellite to quench; the so-called ``quenching time'' \citep[eg.,][]{mcgee_11_short, delucia_infall, wetzel_timescale}. This is interpreted as the length of time after ``accretion'', where accretion is loosely associated with the time a galaxy is within the virialized region of some larger structure, until the galaxy is completely quenched. At low redshift, this timescale is long (several Gyrs) \citep{wetzel_timescale}, leading to the suggestion that gentle stripping of the gas reservoir is the mechanism responsible. However, such a picture naturally leads to most satellite galaxies having relatively paltry star formation rates as their gas is dwindled away. 

In direct contrast, observational studies of satellites have been hampered by the lack of such `smoking guns' of galaxy transformation. Indeed, many recent studies have shown that those satellites which are forming stars are doing so at a rate indistinguishable from central galaxies of similar mass \citep[eg.][]{peng10, mcgee_11_short, wetzel, muzzin_gclass}. To parametrize this behaviour, \citet{wetzel_timescale} introduced the notion of a `delay time' -- that is, the length of time after infall for which a satellite galaxy appears to track the star formation rate of an analogous central galaxy\footnote{The existence of this delay time is still a matter of some debate as several authors have found that star forming galaxies have lower specific star formation rates in satellites than in centrals \citep[eg.][]{patel_2009, vulcani_2010, Haines_2013}. However, some of this discrepancy results from the definition of star forming galaxies, where authors who have more inclusive definitions find the discrepancy. Nonetheless, the existence of a tail of low star formation rate galaxies does not affect the overall observation that most satellites are forming stars at normal rates for a significant time.}.  In context of a rapid baryon cycle, the length of this delay time holds clues for our understanding of galaxy formation as it is expected that the fresh gas supply is also cut off near infall. Dark matter simulations show that the accretion of mass into a subhalo stops when it reaches 1.8 R$_{\mathrm{vir}}$ (as defined in \citep{Bryan_norman_98}) of a larger halo \citep{behroozi_accretion}. This limits the reservoir of gas which can be used for future star formation of the satellite, even in the absence of stripping. 

The final relevant timescale of satellite quenching is a `fading time'. This is equal to the quenching time minus the delay time, and is interpreted physically as the time it takes a satellite to be quenched once its star formation rate deviates from that of an analogous, central galaxy. This phase is short ($<$ 1 Gyr) and is reflected in the relative lack of `green' or intermediate star forming galaxies.

In this letter, we compile measurements of the quenching and delay times of satellites at a range of redshifts to determine if `orbit-based' or `outflow-based' models dominate the quenching. Further, we can directly put limits on the outflow rates allowed by such delay times. We finish by exploring the difficulty of achieving long delay times in standard models of galaxy formation. In this letter, we adopt a \LCDM cosmology with the parameters: $\Omega_{\rm m} = 0.27$, $\Omega_{\Lambda}=0.73$ and $h=H_0/(100 \kmsmpc)=0.70$. All stellar masses assume a \citet{chabrier_imf} initial mass function.

\section{Observed satellite quenching timescales}\label{sec-obsquench}

The best current measurement comes from the analysis of \citet{wetzel_timescale} from a low redshift (z$\sim$0.05) SDSS sample of galaxy groups, where the activity level of the galaxies is determined from a H$\alpha$-based spectroscopic measurement. Wetzel et al. have carefully tracked the infall history of satellites at this epoch, accounting for the evolution in infalling field star formation rates, to find that a satellite galaxy with a stellar mass of 10$^{10.5}$ \Msun has a total quenching time of 4.4 $\pm$ 0.4 Gyrs. They find that 3.8 $\pm$ 0.4 Gyrs of this time is due to the `delay time', while the remaining $\sim$ 0.6 Gyrs of the quenching time is the `fading' stage. \citet{wetzel_timescale} define the moment a galaxy becomes a `satellite' as the time the subhalo crosses 200 times the matter background density (R$_{200,b}$), not the critical density (R$_{200,c}$). This R$_{200,b}$ definition is similar to the R$_{\mathrm{vir}}$ used in \citep{behroozi_accretion}. These densities evolve differently with redshift, so we consistently assume the Wetzel et al. definition and make corrections when the observational results were presented differently. At z=0.0 this matter density radius is 1.92 times larger than the critical density radius, although by z=1.0 it is only 1.15 times larger.

Using data from the GEEC2 sample of spectroscopic galaxy groups at z $\sim$ 0.9, \citet{mok_timescale} presented an estimate of the quenching and delay timescales by examining the relative fractions of `star forming', `intermediate' and `quiescent' satellites as defined by their locations in a colour-colour plane designed to separate the contributions from age and dust. Using an infall model and parametrization of the star formation histories of central galaxies, these authors concluded that these z $\sim$ 0.9 satellites are best fit by a total quenching time of 1.05 $\pm$ 0.25 Gyrs, made up of a 0.8 $pm$ 0.25 Gyr delay time and the remaining  $\sim$ 0.25 Gyr due to the fading time. This timescale applies to 10$^{10.5}$ \Msun\ galaxies, but is not seen to be a strong function of stellar mass for this sample.

From a recent comparison between the phase space distribution of post-starburst galaxies in z $\sim$ 1 galaxy clusters from the GCLASS sample and the phase space of subhalos accreted into cosmological-zoom simulations of galaxy clusters, \citet{muzzin_timescale} find that the total timescale for quenching is $\sim$ 1 $\pm$ 0.25 Gyr. The characteristic phase space features of these post-starburst galaxies can be best represented if quenching occurs quickly (0.1 - 0.5 Gyrs) following the first passage of 0.25 - 0.5 R$_{200,c}$. This is consistent with an analysis of the spectral features of the post-starbursts, selected to have low D$_n$(4000) and a lack of [OII] emission, which suggests that they have a short fading time ($\sim$ 0.1 - 0.5 Gyr). This implies that the delay time is the travel time from R$_{200,b}$ to this position. Directly from this simulation, this delay time is then 0.45$\pm$0.15 Gyrs depending on which inner distance the quenching occurs at. While the Muzzin et al. timescale is not presented in bins of stellar mass, they do state that most of the relevant galaxies are in the range of 10$^{10.5}$ \Mstel, and thus should be comparable to our previous timescales. 

We also add a quenching time derived from the \galex-based star formation rates of galaxy groups at z$\sim$0.4 from \citet{mcgee_11_short}. We can define the excess quenching between groups and the field in a manner similar to \citet{vandenbosch_2008} as $\frac{f_{q,grp} - f_{q,fld}}{1 - f_{q,fld}}$, where $f_{q,grp}$ and $f_{q,fld}$ are the quenched fraction of the groups and the field, respectively. In 10$^{10.5}$\Msun galaxies, this excess quenching is 0.32 $\pm$ 0.1. By comparison with the accretion histories as compiled in \citet{mcgee_accretion_short}, we find that at z=0.4, 30$\%$ of galaxies became satellites more than 4.0$\pm$ 0.6 Gyrs ago. Although these authors do not explicitly measure the fading time, for our purposes we assume it is also 0.5 Gyrs. We caution that even minor differences in the criteria to demarcate between star forming and passive can lead to large differences within different surveys. In the future, a more comprehensive and homogeneous analysis of the existing data will be useful.

\section{Analysis}

\subsection{Orbit-based or outflow-based quenching?}

The time it takes a galaxy to proceed from the virial radius of a galaxy cluster to a specific location (e.g. perigee) in its orbit depends on redshift. Perhaps the simplest hypothesis is that the efficiency of gas stripping depends only on the orbit. That is, the same satellite will be stripped of the same fraction of gas at perigee, in the same orbit, regardless of whether this occurs at $z=0$ or $z=1$. In this case, the quenching timescale of galaxies from `orbit-based' quenching mechanisms will evolve as the ratio of the inverse densities, $\propto$ (1 + z)$^{-3/2}$.\footnote{The crossing time of a cluster is the ratio of the radius, R, and the velocity, V. For a cluster in virial equilibrium the cluster mass, M, scales as RV$^2$ and for a spherical cluster as R$^{3}\rho$. So, the dynamical time, R/V, scales as $\rho^{-1/2}$, while the density, $\rho$, scales as (1+z)$^3$.}

In contrast, an outflow-based quenching mechanism should have a timescale which scales as the star formation rate. Once the cosmological accretion of gas is halted from penetrating into the sphere of influence of a satellite galaxy, the timescale on which the galaxy quenches is equal to the total gas available at this point divided by the `gas consumption' rate. While this gas consumption rate is usually taken to be equal to the star formation rate of the galaxy, this ignores the outflows of gas associated with this star formation.  It is partly for this reason that satellite galaxies in semi-analytic models are quenched more quickly than the star formation rate alone would suggest. In this scenario, satellites prematurely halt their growth because they continue vigorous star formation, and drive associated outflows, despite a significant restriction on their available resources.  We therefore borrow a term from the field of ecology and refer to this quenching mechanism as ``overconsumption''.

For our purposes in this section, the exact outflow rate is not important if it is invariant in redshift for a given star formation rate (SFR) and stellar mass (M$_{stel}$). In that case, any outflow-based quenching would require the quenching time to scale with the star formation rate, or the specific star formation rate (sSFR = SFR/M$_{stel}$). Several authors have presented fitting formula to the observed star forming `sequence' of galaxies over cosmic time. In particular, \citet{peng10}, found that 
\beq
\mathrm{sSFR} = 2.5 \left(\frac{M_{stel}}{10^{10} M_\odot}\right)^\beta \left(\frac{t}{3.5 \mathrm{Gyr}}\right)^{-2.2}
\eeq
where $t$ is the age of the universe and $\beta$ describes the tilt of the specific star formation rate relation, which we take to be 0. Similarly, \citet{whitaker_2012} parametrized the star formation rate as
\beq
\mathrm{log}(\mathrm{SFR}) = \alpha(z)(\mathrm{log} M_{stel} - 10.5) + \beta(z), 
\eeq
where $\alpha(z) = 0.70 - 0.13z$ and $\beta(z) = 0.38 +1.14z - 0.19z^2$. We include these two relations as a measure of the uncertainty in such evolutionary measurements of the sSFR, and note that such relations rely heavily on the assumed functional form at very low and very high redshift where the fit is not well constrained by the data \citep[eg.,][]{Behroozi_2013}.

\begin{figure}
\hspace{-0.7cm}
\leavevmode \epsfysize=6.3cm \epsfbox{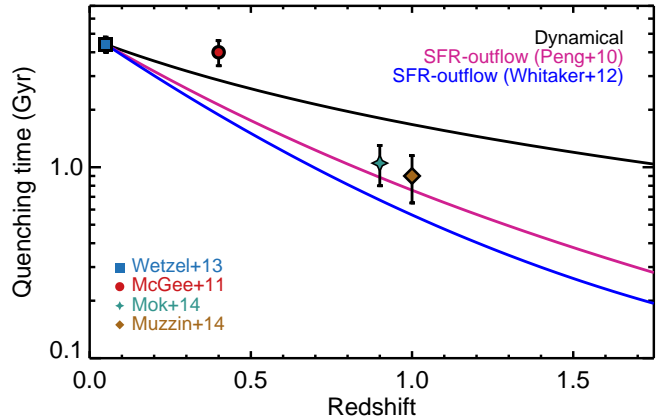}
\caption{The evolution of satellite quenching time with redshift. The lines represent the expectation from an `orbit-based' or `outflow-based' quenching mechanism. The data points represent the observed quenching times of \citet{wetzel_timescale}, \citet{mcgee_11_short}, \citet{mok_timescale} and \citet{muzzin_timescale} as described in \textsection \ref{sec-obsquench}.}
\label{fig_scaling}
\end{figure}

In Figure \ref{fig_scaling}, we show the evolution of quenching time expected from orbit-based and outflow-based quenching (using either the Peng et al. or Whitaker et al. star formation rates with \Mstel\ = 10$^{10.5}$\Mdot) when normalized to a quenching time of 4.4 Gyr at z=0.05. These are shown along with the observed quenching times described in \textsection \ref{sec-obsquench}. While the lower redshift point is roughly consistent with the dynamical time evolution, the z $\sim$ 1 quenching times seem to be lower than expected. This could be due either to an onset of outflow-based quenching or an enhanced stripping efficiency at high redshift. We note that \citet{tinker_wetzel_2010} and \citet{tinker_2013_short} have concluded that their halo model/clustering measurements are consistent with dynamical evolution, although they do not compare with SFR-derived timescales. In the next section, we will determine the outflow rates required to explain these quenching times. 

%%%%%%%%%%%%%%%%%%%%%%%%%%%%%%%%%%%%%%%%%%%%%%%%%%%%%%%%%%%%%%%
%%%%%%%%%%%%%%%%%%%%%%%%%%%%%%%%%%%%%%%%%%%%%%%%%%%%%%%%%%%%%%%%

\subsection{Limits on the outflow rate}

As discussed in the introduction, one of the most curious observed aspects of satellite galaxies is that those which are forming stars appear to have similar star formation rates to central galaxies of similar mass -- requiring a `delay time'. Given the halting of cosmological accretion, the largely unchanging star formation rate during the delay time must result from an unchanged fuelling rate of the cold gas disk from the halo's gas reservoir, and will result in an unchanged outflow rate driven by this star formation. Thus, even if the delay time is established by the onset of orbit-based quenching, there must be a sufficient reservoir at satellite infall to maintain this fuelling throughout. Given the delay time, we can determine the maximum allowed outflow rate from basic baryon accounting measures. 

The inflow and outflow of gas from a galaxy halo can be parametrized as  
\beq
\dotMin + \dotMres = \dotMstar (1 - R) + \dotMout
\label{equilibrium_equation}
\eeq
where $\dot{M}_x$ is the rate of gas accretion due to cosmological infall (x = $in$), the rate of change of the gas reservoir of the system (x = $res$), the star formation rate (x = $stel$) and the rate of outflows from feedback (x = $out$). $R$ is fraction of gas returned to the ISM through supernova, stellar winds, etc., and we assume that this is done instantaneously. This basic equation has been the basis of galaxy formation models for some time \citep[eg.,][]{Colemodel}, and has recently been recast as the `equilibrium condition' by several authors \citep[eg.,][]{Bouche_2010, Dave_2012, Lilly_2013}, although the precise definition varies depending on the definition of the system (eg., as either the galaxy alone, or the galaxy and its extended halo) and whether a `reservoir' exists. Notice that a non-zero delay time immediately suggests that a sizeable reservoir must exist upon satellite infall. 

Assuming that cosmological accretion is halted on infall, then $\dotMin$ = 0. As noted previously, there are simulations which suggest this happens to the dark matter prior to R$_{\mathrm{vir}}$ and, in such cases, we would overestimate the maximum outflow rate. The maximum time that the star formation rate can be maintained at a constant rate is 

\beq
T_{\mathrm{delay}} = \frac{M_{res}}{\dotMres}.
\eeq
Taking account of the possible locations of the halo's baryon content at the point of infall, M$_{res}$ can be given from a conservation equation,
\beq
M_{res} = M_{baryon} - M_{cold} - M_{stel} - M_{ej} - M_{strip},
\label{conservation_equation}
\eeq
where $M_{baryon}$ is the total mass of baryons that {\it belong} to the system, and can be taken as the universal fraction of baryons multiplied by the satellite halo mass ($f_{baryon}*M_{halo}$). $M_{ej}$ is the mass of baryons that have been ejected from this system prior to infall. In our discussion, we will assume that baryons ejected are lost from the system, with no re-accretion. By this definition the integral of $\dotMout$ from its formation time to infall is equal to $M_{ej}$. This is largely just a definition of our terms, and therefore defines the outflow rate to be the rate at which gas is expelled from the system forever. $M_{strip}$ is the mass of gas which could be stripped during the lifetime of the quenching phase, and can be held to zero to find the maximum possible outflow rate. $M_{cold}$ is the cold gas currently locked up in the galaxy. It is likely that this mass must be approximately constant throughout the delay time in order to produce an approximately constant star formation rate according to the Kennicutt-Schmidt law \citep{Schmidt_1959, kennicutt}. Perhaps it is the eventual depletion of this gas that results in the `fading time'.

In the case of outflows produced from star formation, either directly through radiation pressure or from subsequent supernova explosions, the outflow rate is related to the star formation rate and is often parametrized as a mass-loading factor, $\eta$ ($\equiv$ $\dotMout$/$\dotMstar$). Using the equilibrium equation (equation \ref{equilibrium_equation}), along with the conservation equation (equation \ref{conservation_equation}) and the definition of the mass loading factor, the delay time is

\beq
T_{\mathrm{delay}} = \frac{M_{baryon} - M_{cold} - M_{stel} - M_{ej} - M_{strip}}{M_{stel} (1 - R+\eta) sSFR}
\eeq
where we have defined the specific star formation rate (sSFR = $\dotMstar$/M$_{stel}$). For the purposes of parametrizing our knowledge of the baryonic components it is convenient to define each component with respect to the halo mass at infall, so that $f_x$ is $M_x$/$M_{halo}$. $R$, the returned gas fraction, is usually on the order of $\sim$ 0.25-0.4, depending on the choice of initial mass function and the time since the stellar population was born. We assume a value relevant for a Chabrier IMF with a 2 Gyr population (R = 0.4).

We can now estimate each of these components. In our assumed cosmology, $f_{baryon}$ = 0.17. There have been many attempts to quantify $f_{stel}$, largely through halo abundance matching techniques. It has been shown that this fraction is constant with redshift, assuming a universal initial mass function, although it does strongly depend on halo mass \citep[eg.,][]{moster_hod, guo_2011_efficiency, Behroozi_2013, mcgee_imf}. For our canonical 10$^{10.5}$\Msun\ stellar mass galaxy, using the functional form presented by \citet{moster_hod}, we find that this stellar fraction is 0.03. Although we adopt $f_{stel}$ =0.03, it is worth noting that \citep{Behroozi_2013} find similar stellar fractions in the range 0.024 to 0.028 to z$\sim$ 3 at this stellar mass. \citet{carilli_walter_2013} have compiled existing measurements of the H$_{2}$ content of strongly star forming galaxies at a range of redshifts and found that $f_{cold}$ = 0.1 * (1+z)$^2$ $f_{stel}$. As the H$_{2}$ is expected to correlate well with the star formation rate, this is an appropriate value for the cold gas content, rather than the sum of molecular and atomic hydrogen gas. In our parametrization atomic hydrogen will be part of the gas reservoir. If we assume that the ejected gas is a result of past outflows from star formation rates, then $f_{ej}$ = $\eta$ $f_{stel}$ (1+$R$), where the $R$ is needed to recover the total past star formation rate. This model of past gas ejection is clearly an approximation, as earlier star formation occurred as the halo was forming and thus $\eta$ is unlikely to be constant throughout. Indeed, given the lower potential wells, and the need to widely eject metals into the intergalactic medium to match observation, this $\eta$ may be an underestimate of the amount of ejected material. As well, this assumes that the star formation--driven outflows dominate the ejected gas budget. 

Using these approximations, the delay time is 

\beq
 T_{\mathrm{delay}} = \frac{f_{baryon} - f_{stel}(0.1*(1+z)^2 + \eta (1+R) + 1) - f_{strip}}{f_{stel} (1-R+\eta) sSFR}
\label{equ-delay}
\eeq

\begin{figure}
\hspace{-0.7cm}
\leavevmode \epsfysize=6.0cm \epsfbox{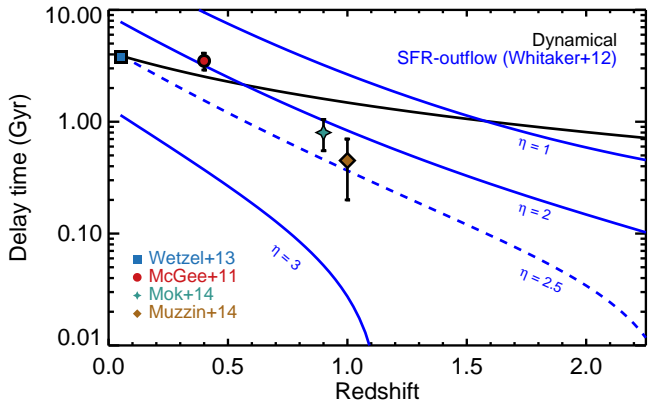}
\caption{The maximum possible delay times inferred as a function of redshift for a given mass loading factor ($\eta$ = 1, 2, 2.5, 3) and assuming a \citet{whitaker_2012} star formation history of star forming galaxies. The dynamical time scale is also shown. }
\label{fig_delay}
\end{figure}

We note that Equation \ref{equ-delay} has only one free parameter, $\eta$, and so can be completely characterized by a measure of the satellite delay time. In Figure \ref{fig_delay}, we show the maximum delay times (assuming $f_{strip}$ = 0) as a function of redshift for $\eta$ = 1, 2, 2.5, 3. We have also plotted the delay times from the observations, and a dynamical scaling which fits the delay time at z=0.05. Notice that the scaling with any particular $\eta$ is much steeper than the dynamical scaling because the evolution of the observed sSFR is very rapid. The turnover in delay time at $\eta$ = 3 and z $>$ 1.0, or $\eta$ = 2.5 and z $>$ 2.0, results from the need to maintain ever larger fractions of cold gas in the disk, while past outflows caused by the high $\eta$ make this impossible. The incompatible slope of the star formation rate and that implied by the dynamical time is a long-standing problem for models, and may suggest an underlying change in the efficiency of star formation, either through varying wind re-accretion or small scale cloud conditions.

Interestingly, in Figure \ref{fig_delay}, we see that the z $\sim$ 1 observations are consistent with a maximum mass loading of 2.5. Observations of strongly star forming galaxies suggest that the mass-loading is on the order of unity \citep{Heckman_2000, martin_outflows_2005}, although the measurements are difficult to interpret. While the ultimate fate of this outflowing gas is unclear, some simulations suggest that up to a third of the gas is re-accreted \citep{oppenheimer_dave_2008}, although this does not necessarily apply in satellites. In semi-analytic models, which are designed to reproduce the luminosity/mass function, the mass loading factors are typically much greater than one, and are a strong function of halo mass \citep[eg.,][]{guo_2011, Bower2012}. Although much of this gas may also be re-accreted, it is again not clear if this is the case in satellites \citep{font}. However, notice that with even moderate mass loading of 1, the expected delay time is shorter than expected from a scaling of the dynamical time in z $>$ 1.5 satellite galaxies. This is a potentially new quenching mechanism for satellite galaxies, and calls for a systematic observational study of such galaxies.

%%%%%%%%%%%%%%%%%%%%%%%%%%%%%%%%%%%%%%%%%%%%%%%%%%%%%%%%%%%%%%%
%%%%%%%%%%%%%%%%%%%%%%%%%%%%%%%%%%%%%%%%%%%%%%%%%%%%%%%%%%%%%%%%

\subsection{Implications for the cooling model}

In the previous section, we have implicitly assumed that the only necessary condition to sustain star formation at a constant rate is sufficient gas in the reservoir. However, such a constant SFR also requires that the cooling rate of gas from the reservoir is unchanged throughout the delay time. In a common model of cooling gas, the bulk of the gas reservoir is in the form of a hot halo, with a temperature that is a result of shock heating to the virial temperature upon accretion. This gas cools, largely through bremsstrahlung, and the rate depends on the square of the gas density. In simple models, as the gas near the central, dense region cools, the gas will respond to the flow quickly, so that the gas density profile shape will be driven by the dark matter potential. This seems like the best case scenario, as it is expected that the hot gas may build up outside the cooling radius \citep{fabian_1977}.) Thus, while the profile shape is constant, the decline in the total gas content will cause the central density to decline. The cooling rate is then directly linked to the amount of the gas in the reservoir. As the hot gas reservoir begins to drain, the cooling rate will go down, the subsequent amount of cold gas will decline and thus, star formation will also decline.

One solution to maintain uniform cooling rates is to change the shape of the gas profile as it is drained. It is possible that the immersion of a satellite galaxy into the hot, dense intracluster medium will compress the hot halo. However, given the need to match the compression rate with the cooling rate regardless of the exact density of the ICM, this suggests a fine tuning problem. Unfortunately, it is very difficult to observe X-ray halos of galaxies, and thus to determine this directly. One alternative to such a fine tuning is that the accreted gas is never shock heated to the virial temperature, and simply adds to the atomic gas reservoir. The thermal history of gas in galaxies has long been studied \citep{white_rees}, and is still debated \citep[eg.,][]{keres_2005, Dekel_birnboim}. Interestingly, though, this may hint that the cooling rate is governed by the interplay between atomic and molecular gas rather then the hot gas cooling. As we have mentioned, our definition of the gas `reservoir' includes the HI content of gas. In principle, if this is the primary gas reservoir, we would expect to see a decrease in the HI content for satellite galaxies which are strongly star forming in relation to that seen in central galaxies. In a future paper, we will directly measure the HI content of star forming satellites for indications that this is being drained, and thus place stronger constraints on the cooling mode.

\section{Conclusions}
We have compiled measurements of the quenching and delay times of satellite galaxies as a function of redshift from the literature. We have used these measurements, in combination with a simple model for the baryon cycle in galaxies, to arrive at the following conclusions:

\begin{itemize}
\item While our compilation of measurements may be subject to systematic effects, the evolution of quenching timescales is consistent with the scaling expected if the quenching was driven not by `orbit-based' stripping events like ram pressure stripping or strangulation but rather by secular outflows in light of a halt of cosmological accretion (overconsumption).  Alternatively, it may be that `orbit-based' stripping is significantly more effective at high redshift than a dynamical scaling would suggest. 

\item The delay time of satellite galaxies places an upper limit on the rate of outflow driven gas, or the mass-loading, which from current data is $\eta$ = 2.5 in \Mstel\ = 10$^{10.5}$\Msun\ galaxies. Further measurements at higher redshift and lower masses will put even tighter constraints on the outflow rate. 
 
\item The delay time requires an uninterrupted rate of cooling gas from the gas reservoir. This presents a challenge to typical models of cooling gas for which the cooling rate is strongly dependent on the density of the reservoir, and thus declines quickly as the reservoir drains. 

\item We predict that at z$>$1.5 satellite galaxies are quenched through secular outflow processes instead of external stripping events. Future observations of such galaxies will put constraints on the outflow rates of all star forming galaxies and probe this unique quenching mechanism.

\end{itemize}
\section*{Acknowledgments}
SLM acknowledges support from an NWO grant to George Miley. RGB is supported by STFC grant ST/I001573/1. MLB acknowledges support from an NSERC Discovery grant, and from NOVA and NWO visitor grants that supported his sabbatical visit to Leiden Observatory, where this work was undertaken.

\bibliography{ms}

\end{document}